\documentclass[conference,10pt,a4paper]{IEEEtran}

\usepackage[T1]{fontenc}
\usepackage{url}
\usepackage{cite,xcolor}
\usepackage[cmex10]{amsmath} 

\usepackage[cmintegrals]{newtxmath}
\usepackage{bm} 
\usepackage{graphicx}
\usepackage{enumitem}
\usepackage{soul}
\setstcolor{purple}
\usepackage{microtype}
\usepackage{algorithm, tabularx}
\usepackage{algpseudocode}
\usepackage{svg}
\usepackage{subcaption}

\graphicspath{{.}{../pictures/}{pictures/}{../../pictures/}}

\definecolor{awesome}{rgb}{1.0, 0.13, 0.32}

\newcommand{\code}[1]{\texttt{#1}}

\newtheorem{lemma}{Lemma}
\newtheorem{example}{Example}
\newtheorem{remark}{Remark}

\newtheorem{proposition}{Proposition}

\newtheorem{definition}{Definition}

\usepackage{etoolbox}
\makeatletter
\patchcmd{\@begintheorem}{\textit}{\textbf}{}{}
\patchcmd{\@begintheorem}{\itshape}{\bfseries}{}{}
\makeatother

\def\gap{1.3ex}
\abovedisplayskip\gap
\belowdisplayskip\gap
\abovedisplayshortskip\gap
\belowdisplayshortskip\gap

\begin{document}
\title{Preferential Pliable Index Coding} 

\author{\IEEEauthorblockN{Daniel Byrne and Lawrence Ong}\thanks{This research was supported by the Australian Research Council (under projects FT140101368 and FT190100429).}
\IEEEauthorblockA{The University of Newcastle}
\and
\IEEEauthorblockN{Parastoo Sadeghi}
\IEEEauthorblockA{University of New South Wales}
\and
\IEEEauthorblockN{Badri N.\ Vellambi}
\IEEEauthorblockA{University of Cincinnati}
}

    

\IEEEoverridecommandlockouts
\IEEEaftertitletext{\vspace{-1\baselineskip}}

\maketitle

\begin{abstract} 
We propose and study a variant of pliable index coding (PICOD) where receivers have preferences for their unknown messages and give each unknown message a preference ranking. We call this the preferential pliable index-coding (PPICOD) problem and study the Pareto trade-off between the code length and overall satisfaction metric among all receivers. We derive theoretical characteristics of the PPICOD problem in terms of interactions between achievable code length and satisfaction metric. We also conceptually characterise two methods for computation of the Pareto boundary of the set of all achievable code length-satisfaction pairs. As for a coding scheme, we extend the Greedy Cover Algorithm for PICOD by Brahma and Fragouli, 2015, to balance the number of satisfied receivers and average satisfaction metric in each iteration. We present numerical results which show the efficacy of our proposed algorithm in approaching the Pareto boundary, found via brute-force computation.

\end{abstract}

\section{Introduction}

In traditional communication scenarios, receivers request specific messages from the sender. Pliable communications, on the other hand, is a bandwidth-efficient communication approach, where instead of sending specific messages to the receivers, it suffices to convey messages that the receivers do not already have. Such flexible transmission finds applications in streaming, key distribution, and distributed learning.

So far, studies on pliable communications have been focusing on the extreme opposite of traditional communications in the sense that each receiver can be equally satisfied with \textit{any} message it does not have. In this paper, we consider a more realistic formulation where not all messages are equally desirable to the receivers. Our objective is to find codes that achieve both good communication rate and good \textit{overall satisfaction metric} that incorporates receivers' preferences.

We study such pliable communications with receiver preferences in the framework of index coding~\cite{baryossefbirk11,lubertzkystav09,neelytehranizhang13,arbabjolfaeikim18trends,ongholim16,ong2017}: a noiseless broadcast network with one sender and multiple receivers. Under the assumption that each receiver already possesses some messages a priori as side information, this seemingly simple framework is sufficiently general to capture the general network-coding formulation~\cite{rouayhebsprintsongeorghiades10,effrosrouayheblangberg15}.

Pliable communication without receiver preferences has also been studied under the framework of index coding, and is called~\emph{pliable index coding}~\cite{brahmafragouli15}. Linear codes have been constructed using iterative heuristic algorithms~\cite{brahmafragouli15,songfragouli18,songfragoulizhao20} and graph colouring~\cite{Krishnan2021,BSubramanianKrishnan22}. The optimal code length was obtained for special receiver side-information configurations~\cite{liutuninetti20,ongvellambikliewer2019,OngConf3}. 


Through fixing the message to be delivered to each receiver (i.e., the \textit{decoding choice} of the receiver), pliable index coding reduces to index coding.  A brute-force optimal solution for pliable index coding can, therefore, be obtained by comparing index-coding solutions over all possible decoding choices and choosing one that achieves the minimum code length.

Heuristic algorithms for pliable index coding avoid this search with exponentially high complexity by iteratively finding sub-codes that \textit{satisfy} a maximal number of receivers in each iteration, and terminate when all receivers are satisfied~\cite{brahmafragouli15}. 

When the receivers have preferences over the messages they do not have---we formulate this problem in this paper and call it \textit{preferential pliable index coding (PPICOD)}---instead of finding the shortest code length, the aim is to find the Pareto boundary of the bi-objective problem of code length and receivers' overall satisfaction metric.

Pliable communications with receiver preferences have been investigated~\cite{songfragouli18b}, where the sole objective was to maximise the overall satisfaction. Heuristic algorithms were proposed to construct codes with a pre-specified length, where in each iteration, one linear combination of messages that maximise the satisfaction of the receivers is constructed, and receiver side information is updated.

Following the previous argument, the Pareto boundary for PPICOD can be obtained by brute force by considering all possible decoding choices (each giving an overall satisfaction metric) and finding the shortest code for each overall satisfaction metric (by solving the corresponding set of index-coding problems). However, this is still prohibitively complex.

In this paper, we propose a heuristic algorithm with a different perspective. In each iteration, instead of focusing only on either the number of receivers satisfied or the satisfaction metric of those receivers,
we devise an objective function that carefully balances both metrics. We numerically demonstrate that this algorithm can be tuned to achieve a trade-off between the code length and the  overall satisfaction metric, and can perform close to the Pareto boundary. We also theoretically detail general characteristics of the PPICOD problem.



\section{Formulation PPICOD}\label{sec:formulation}
We use the following notation: 
$\mathbb{Z}^+$ denotes the set of natural numbers, 
$[a:b] := \{a, a+1, \dotsc, b\}$ for $a,b\in\mathbb{Z}^+$ such that $a < b$, 
$X_S = (X_i: i \in S)$ for some ordered set $S$, 
and
$\mathbb{R}^+$ denotes the set of positive real numbers.

Consider a sender with $m \in \mathbb{Z}^+$ messages, denoted by $X_{[1 : m]} = (X_1, \dots, X_m)$. Each message $X_i$ takes a value in a finite field $\mathbb{F}_q$ of size $q \in \mathbb{Z}^+$. There are $n$ receivers, and receiver~$i \in [1:n]$ already knows a subset of messages $X_{H_i}$, for some $H_i \subsetneq [1:m]$. The set $H_i$ is called the receiver's side information.

In pliable index coding, each receiver is satisfied if it decodes any message that it does not have, that is, receiver~$i$ must decode $X_j$ for some $j \in [1:m] \setminus H_i$. For preferential pliable index coding (PPICOD), the messages $X_{[1:m]\setminus H_i}$ are not equally desired by receiver~$i$, and so each message is given a preference rank. The side information and the message ranking of all receivers are defined using a \textit{preference matrix} $\boldsymbol{P}$:

\begin{definition}
The preference matrix $\boldsymbol{P}$ of a PPICOD problem is an $n\times m$ matrix of the following form:
\begin{equation}
    \boldsymbol{P} = [P_{i,j}]_{n\times m},
\end{equation}
where $P_{i,j}\in\mathbb{R}\cup\{\infty\}$ is receiver~$i$'s preference rank for message $X_j$.
If $j\in [1:m] \setminus H_i$, this rank is finite; else, it is infinite. The $i$-th row of $\boldsymbol{P}$ is referred to as the preference vector of receiver~$i$.
\end{definition}

\begin{remark}
   Receivers give their more preferred messages a lower preference rank, i.e., decoding message~$k$ over message~$l$ with $P_{i,k} < P_{i,l}$ results in a \emph{better} satisfaction for receiver~$i$. 
\end{remark}

A PPICOD problem is fully characterized by $\boldsymbol{P}$. Given $\boldsymbol{P}$, a code consists of
\begin{itemize}
\item an encoding function of the sender, $\mathsf E: \mathbb{F}_q^m \rightarrow \mathbb{F}_q^\ell$,
\item a decoding choice, $D: [1:n] \rightarrow [1:m]$,
\item for each receiver~$i \in [1:n]$,
a decoding function\\ $\mathsf G_i: \mathbb{F}_q^\ell \times \mathbb{F}_q^{|H_i|} \rightarrow \mathbb{F}_q$,
\end{itemize}
such that
\vspace{-2ex}
\begin{align}
    & D(i) \in [1:m] \setminus H_i,\\
    & \mathsf G_i(\mathsf{E}(X_{[1:m]}),X_{H_i}) = X_{D(i)}.
\end{align}
for all $i \in [1:n]$ and for all $X_{[1:m]} \in \mathbb{F}_q^m$.
$D(i)$ is the index of the message decoded by receiver~$i$, and $\ell$ is the code length.

In this paper, we consider only linear encoding functions, that is, $\mathsf{E}(X_{[1:m]}) = \boldsymbol{A} \boldsymbol{X}$, for some matrix $\boldsymbol{A} \in \mathbb{F}_q^{\ell \times m}$ where $\boldsymbol{X} = [X_1\; X_2\; \dotsc\; X_m]^\text{T}$ is a vector consisting of the messages.

A code length--satisfaction pair $(\ell,s)$ is achievable if there is a code with code length $\ell$ and \emph{overall} satisfaction metric
\vspace{-1ex}
\begin{equation}
        s = \sum_{i=1}^n P_{i,D(i)}. \label{eq:global-satisfaction}
    \end{equation}
Better overall satisfaction among receivers means lower $s$. We drop the term ``overall'' when the context is clear.\footnote{Equation \eqref{eq:global-satisfaction} considers the \textit{linear sum} satisfaction metric. Fairness can be considered by changing it to $\max_i P_{i,D(i)}$.} A code length $\ell$ is achievable if $(\ell,s)$ is achievable for some $s$. Similarly, a satisfaction metric $s$ is achievable if $(\ell,s)$ is achievable for some code length $\ell$. For two points $(\ell_1,s_1)$ and $(\ell_2,s_2)$, $(\ell_1,s_1)$ is said to \textit{dominate} $(\ell_2,s_2)$ if and only if 
\begin{equation}
    \ell_1 \leq \ell_2 \quad \text{ and } \quad s_1 \leq s_2,
\end{equation}
with at least one of the inequalities being strict.
For a set of achievable code length--satisfaction pairs $\mathcal{T}$, the Pareto boundary, denoted by $\mathcal{T}^*$, is the set of all the non-dominated points in $\mathcal{T}$.
Define the feasible region $\mathcal{R}$ as the union of achievable $(\ell,s)$ pairs over all linear codes.
 We aim to find the Pareto boundary of $\mathcal{R}$, denoted by $\mathcal{R}^*$. Define $\ell^*$ to be the minimum achievable code length over $\mathcal{R}^*$, and $s^*$ to be the minimum achievable satisfaction metric over $\mathcal{R}^*$.

In this paper, we consider the case where each receiver needs to decode only one message (this is denoted by $t=1$ in the literature~\cite{brahmafragouli15,liutuninetti20}).

\begin{remark}
The formulation of PPICOD specializes to
\begin{enumerate}
    \item pliable index coding by setting $P_{i,j} = 1$ for all $i \in [1:n]$ and all $j \in [1:m] \setminus H_i$. The optimal pliable index code length is then $\ell^*$.
    \item index coding by setting $P_{i,d_i} = 1$ for each receiver~$i$ with a uniquely wanted message whose index is $d_i \in [1:m] \setminus H_i$, and $P_{i,j} = n+1$ for all $j \in [1:m] \setminus (H_i \cup d_i)$. The optimal index code length is then the minimum $\ell$ in $\mathcal{R}^*$ for which the satisfaction metric is $s=n$. Alternatively, we can just restrict the codes such that the decoding choice must be $D(i) = d_i$ and find the minimum $\ell$ in $\mathcal{R}$.
\end{enumerate}
\end{remark}

\section{Deriving Characteristics of PPICOD}
In this section, we derive theoretical characteristics of PPICOD.
\begin{proposition}\label{rate-larger-than}
If $(\ell,s)$ is achievable, then $(\ell',s)$ for all $\ell' > \ell$ are also achievable.
\end{proposition}

\begin{IEEEproof}
The code that achieves $(\ell',s)$ can be constructed by padding the encoding function for the code that achieves $(\ell,s)$ with $\ell'-\ell$ zeros.
\end{IEEEproof}

\begin{proposition}\label{prop:other-s-maybe-unachievable}
That $(\ell,s)$ is achievable does not imply that $(\ell,s')$ for any $s' \neq s$ is achievable.
\end{proposition}

\begin{IEEEproof}
Consider a problem instance with
\begin{equation}
    \boldsymbol{P} = 
    \begin{bmatrix}
    2 & \infty & 1 &\infty & 2\\
    \infty & 1 & 2 & 1 & \infty \label{eq:ppicod-instance}
    \end{bmatrix}.
\end{equation}
The code with $\mathsf E(X_{[1:3]}) = X_3$ and $D(1) = D(2) = 3$ achieves the code length--satisfaction pair $(1,3)$. In order to achieve $s=2$, we must have 
$D(1)=3$, and $D(2)=2$ or $4$.
For $s=4$, the decoding choice must be $D(1)=1$ or $5$, and $D(2)=3$. For all these decoding choices, the problem can be converted into index-coding instances, and then we can show that $\ell \geq 2$. Therefore, neither $(1,2)$ nor $(1,4)$ is achievable.
\end{IEEEproof}

\begin{proposition}\label{rate-cap}
If a satisfaction metric $s$ is achievable, then the code length--satisfaction pair
\begin{itemize}
    \item $(\min\{ n, m - \min_i |H_i| \} ,s)$ is achievable for sufficiently large $q$, and
    \item $(\min\{n,m\},s)$ is achievable for all $q$.
\end{itemize}
\end{proposition}
The proof is provided in Appendix \ref{sec:proof:rate-cap}. Propositions~\ref{rate-larger-than} and \ref{rate-cap} imply that we only need to consider $(\ell,s) \in \mathcal{R}$ where $\ell \leq \min \{ n, m\}$. For sufficiently large $q$, we need to consider only $\ell \leq \min \{ n, m - \min_i |H_i|\}$.

\begin{figure}[t]
    \centering
    \includegraphics[width=0.3\linewidth]{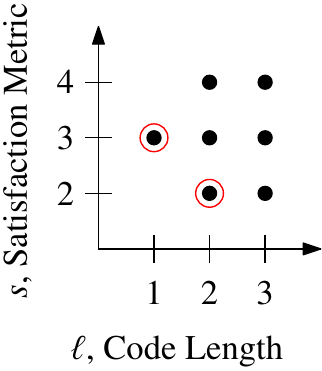}
    \caption{Set $\mathcal{R}$ for the PPICOD problem~\eqref{eq:ppicod-instance}. Black dots are achievable points, and red circles are the Pareto boundary.}
    \label{fig:trade-off}
\end{figure}

\begin{proposition}\label{prop:trade-off}
    For two points $(\ell_1, s_1), (\ell_2, s_2) \in \mathcal{R}^*$, if $s_2 > s_1$, then $\ell_2 < \ell_1$.
\end{proposition}
    
\begin{IEEEproof}
    Suppose that $\ell_1 \leq \ell_2$. Then, $(\ell_1,s_1)$ dominates $(\ell_2,s_2)$. This contradicts that $(\ell_2, s_2) \in \mathcal{R}^*$.
\end{IEEEproof}

Proposition~\ref{prop:trade-off} implies a trade-off between $s$ and $\ell$ along $\mathcal{R}^*$, except for the special case where $\ell^*$ and $s^*$ are simultaneously achieved. This trade-off is true despite the observation in Proposition~\ref{prop:other-s-maybe-unachievable}. Consider the PPICOD problem~\eqref{eq:ppicod-instance}. We can show the following:
    \begin{itemize}
        \item For $s=2$, the minimum $\ell$ is 2, giving $(2,2) \in \mathcal{R}$.
        \item For $s=3$, the minimum $\ell$ is 1, giving $(1,3) \in \mathcal{R}$.
        \item For $s=4$, the minimum $\ell$ is 2, giving $(2,4) \in \mathcal{R}$.
    \end{itemize}
    These three points, along with other achievable points, are plotted in Figure~\ref{fig:trade-off}. The sequence of Pareto boundary points maps out a trade-off between $\ell$ and $s$. Note that the sequence of points along the Pareto boundary need not be convex.

\subsection{Bipartite Graph Representation}

A PPICOD problem can be completely described by a weighted bipartite graph $G=(\mathbb{U}, [1:m], E, w)$, where the receiver set $\mathbb{U} := \{r_i: i \in [1:n] \}$ and the message set $[1:m]$ form the two disjoint and independent sets. Here, node $r_i$ corresponds to receiver~$i$. The set of edges is 
\begin{equation}
 E = \{(r_u,v) : u \in [1:n], v \in [1:m] \setminus H_u\},
\end{equation}
and the weight function $w: E \rightarrow \mathbb{R}^+$ is defined such that each $(r_u,v) \in E$ has weight  $w(r_u,v) = P_{u,v}$. We call such a graph a PPICOD graph.

\begin{figure}[t]
    \centering
    \includegraphics[width=0.3\linewidth]{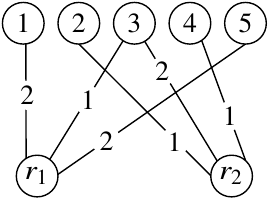}
    \caption{The bipartite graph representation for the PPICOD problem~\eqref{eq:ppicod-instance}. The top row represents the messages, and the bottom row, the receivers. The weight on an edge indicates the receiver's preference ranks for the corresponding message.}
    \label{fig:bipartite-graph}
\end{figure}

\begin{example}
Figure~\ref{fig:bipartite-graph} depicts the PPICOD graph for \eqref{eq:ppicod-instance}.
\end{example}

\subsection{Calculation of the Pareto Boundary} \label{sec:pareto}

\subsubsection{Method 1 (Decoding centric)} It has been pointed out~\cite{liutuninetti20} that by restricting the decoding choice $D$, the problem of finding the minimum code length $\ell$ is an index coding problem. And so, finding the set $\mathcal{R}^*$ can be done in the following way:
\begin{itemize}
    \item[a)] Determine the satisfaction metrics, according to \eqref{eq:global-satisfaction}, for all possible decoding choices.
    \item[b)] For each satisfaction metric $s$ found above, find the minimum code length of each associated index coding problem (each corresponds to one decoding choice for PPICOD that gives the satisfaction metric $s$).
    \item[c)] Record the code length--satisfaction pairs from Step~b.
    \item[d)] Repeat Steps~b and c above for all $s$. Let the collection of all code length--satisfaction pairs be $\mathcal{R}_1$.
    \item[e)] Find the Pareto boundary $\mathcal{R}_1^*$ of $\mathcal{R}_1$.
\end{itemize}

\begin{proposition} \label{prop:r1*=r*}
    $\mathcal{R}_1^* = \mathcal{R}^*$.
\end{proposition}


Refer to Appendix~\ref{app:r1*=r*} for the proof of Proposition~\ref{prop:r1*=r*}. The proof uses the following lemma:
\begin{lemma} \label{lemma:subset-pareto}
If $\mathcal{R}^* \subseteq \mathcal{R}_1 \subseteq \mathcal{R}$, then $\mathcal{R}_1^* = \mathcal{R}^*$.
\end{lemma}
See Appendix~\ref{app:subset-pareto} for the proof of Lemma~\ref{lemma:subset-pareto}.

The complexity of Method~1 is high as it cycles through all possible $\Pi_{i \in [1:n]} (m - |H_i|)$ decoding choices. And for each decoding choice (that is, an index coding problem), there is no efficient way to determine the minimum code length in general. Restricting the code to be linear, however, the minimum code length for each decoding choice can be systematically determined by finding the minimum rank of all matrices that ``fit'' the index-coding problem, commonly referred to as the minrank of the problem~\cite{baryossefbirk11}. Finding the minrank involves calculating the $\mathbb{F}_q$-rank of $q^{\sum_{i \in [1:n]}|H_i|}$ matrices. Using minrank, the complexity of Steps~a and b in Method~1 is $\mathcal{O}(n m^n q^{nm})$.

\subsubsection{Method 2 (Code centric)} Method~1  exploits a common view that pliable index coding is a set of index-coding instances with different decoding choices. Instead, an alternative to searching over decoding choices is a code-centric search, which considers all possible linear codes:
\begin{itemize}
    \item[a)] List all $m \times m$ matrices over $\mathbb{F}_q$ in reduced row echelon form.
    \item [b)] For each matrix $\boldsymbol{A}$ in Step~a, perform the following for each receiver~$i \in [1:n]$. 
    \begin{itemize}
        \item[b1)] Remove columns $H_i$ from $\boldsymbol{A}$ to get $\boldsymbol{A}_{\setminus H_i}$.\footnote{For any matrix $\boldsymbol{A}$, $\boldsymbol{A}_{\setminus H(i)}$ is defined as the submatrix of $\boldsymbol{A}$ with only columns $H(i)$ removed.}
        \item[b2)] Perform Gaussian elimination on $\boldsymbol{A}_{\setminus H_i}$ to obtain the reduced row echelon form of $\boldsymbol{A}_{\setminus H_i}$.
        \item[b3)] The rows that contain only one non-zero element indicate the message that receiver~$i$ can decode. The message index is the column number in $\boldsymbol{A}$.
        \item[b4)] Find the smallest preference rank $P_{i,j}$ among the decodable messages.
    \end{itemize}
    \item[c)] The pair $(\ell,s)$ is achievable, where $\ell$ is the number of non-all-zero rows in $\boldsymbol{A}$ and the overall satisfaction metric $s$ is the sum of the smallest preference ranks of decodable messages from all the receivers. Let the collection of all code length--satisfaction pairs be $\mathcal{R}_2$.
    \item[d)] Find the Pareto boundary $\mathcal{R}_2^*$ of $\mathcal{R}_2$.
\end{itemize}

\begin{proposition}\label{prop:linear:pareto}
    $\mathcal{R}_2^* = \mathcal{R}^*$.
\end{proposition}

Refer to Appendix~\ref{app:linear:pareto} for the proof of Proposition~\ref{prop:linear:pareto}. 
The number of matrices in Step~a is
$\sum_{k=1}^m \binom{m}{k}_q = \mathcal{O}(q^{(m+1)^2/4})$.
Step~b has polynomial complexity in $n$ and $m$. 

We conclude this section by remarking that $\mathcal{R}_1 \neq \mathcal{R}_2$ in general. Although this paper considers linear encoding,  Propositions~\ref{rate-larger-than}--\ref{prop:r1*=r*} also hold in general without this restriction.

\section{An Algorithm for PPICOD}

\subsection{Review of  Greedy Cover Algorithm for Pliable Index Coding}

The Greedy Cover (GrCov) algorithm~\cite{brahmafragouli15} devises codes for pliable index coding problems by iteratively selecting message index subsets to satisfy the receivers.
The algorithm is based on the bipartite graph representation of the problem. Recall that for pliable index coding, all edges have weight one.
In each iteration, a set of message nodes $\mathcal{S}$ and the set of receiver nodes $W_1'(\mathcal{S})$ that $\mathcal{S}$ \textit{satisfies} are determined. The set of receiver nodes that is satisfied by $\mathcal{S}$ must each have only one edge to $\mathcal{S}$, so that the sub-code $\sum_{i \in \mathcal{S}}X_i$ generated by $\mathcal{S}$ allows each of these receivers to decode a new message. 
Formally,
\begin{equation}
    W_1'(\mathcal{S}) := \big\{ r_u \in \mathbb{U}: \big| \{(r_u,v) \in E: v \in \mathcal{S} \} \big| = 1  \big\}. \label{eq:neigbour}
\end{equation}

In each iteration, a maximal set $\mathcal{S}$ is constructed by greedily adding message nodes to an empty set, where the choice of each message node to be added is one that increases $|W_1'(\mathcal{S})|$ the most. After this iteration, $W_1'(\mathcal{S})$ and all edges connecting to them are removed from the graph (recall $t=1$). This process is repeated until all receiver nodes are removed. A code is then formed by concatenating sub-codes $\sum_{i \in \mathcal{S}}X_i$ from all iterations.


\subsection{The Preferential Greedy Cover (PrGrCov): A Modified Algorithm for PPICOD}


\begin{algorithm}[t]
\caption{PrGrCov($G,\alpha,\boldsymbol{\eta}$)}\label{alg:pr-grcov}
\begin{algorithmic}[1]
\State Inputs: A PPICOD graph $G=(\mathbb{U}, [1:m], E, w)$ and parameters $\alpha$, $\boldsymbol{\eta}$
\State $\mathcal{C} \gets \emptyset$
\State $\mathsf{SAT} \gets \emptyset$ 
\While{$|\mathsf{SAT}| \neq n$}
    \State $\textrm{maximal} \gets \code{False}$
    \State $\mathcal{S} \gets \emptyset$
    \While{$\textrm{maximal} = \code{False}$}
        \State $\textrm{maximal} = \code{True}$
        \State Randomly pick $j^* \in \arg \max_{j \in [1:m] \setminus \mathcal{S}} f(\mathcal{S} \cup \{j\})$
        \If{$f(\mathcal{S} \cup \{j^*\}) > f(\mathcal{S})$}
            \State $\textrm{maximal} \gets \code{False}$
            \State $\mathcal{S} \gets \mathcal{S} \cup \{j^*\}$
        \EndIf
    \EndWhile
    \State $\mathsf{SAT} \gets \mathsf{SAT} \cup W_1(\mathcal{S})$
    \State $\mathcal{C} \gets \mathcal{C} \cup \{\mathcal{S}\}$
    \State $E \gets E \setminus \{ (u,v) \in E: u \in W_1(\mathcal{S})\}$
\EndWhile
\State Output: $\mathcal{C}$
\end{algorithmic}
\end{algorithm}

We modify GrCov in the following way for PPICOD. The main idea is to balance the code length~$\ell$ and the satisfaction metric~$s$. 
To this end, instead of maximizing $|W_1'(\mathcal{S})|$ in each iteration as in GrCov, we introduce the following fitness function, for some $\boldsymbol{\eta} := [ \eta_1\; \eta_2\; \dotsm \eta_n] \in (\mathbb{R}^+)^n$ and $\alpha \in [0,1]$:
\begin{align}\label{eqn:fitness-func-improved}
    &f(\mathcal{S}) = 
    \begin{cases}
    -(\eta_{\textrm{max}}+1), & \text{if }|W_1(\mathcal{S})| =0 \\
        \alpha |W_1(\mathcal{S})| - (1 - \alpha) \frac{M (\mathcal{S}) } { |W_1(\mathcal{S})| }, & \text{otherwise}.
    \end{cases}  
\end{align}
We now explain the above function:
\begin{enumerate}
    \item The set of satisfied receivers \eqref{eq:neigbour} is altered to include only edges from each receiver node~$r_u$ of weight at most $\eta_u$:
\begin{align*}
        W_1(\mathcal{S}) := \left\{  
        r_u \in \mathbb{U}:  \begin{matrix}
        \big| \{(r_u,v) \in E: v \in \mathcal{S} \} \big| = 1, \\ 
        \max\limits_{ (r_u,v) \in E : v \in \mathcal{S}} w(r_u,v) \leq \eta_u
       \end{matrix}
        \right\},
\end{align*}
where to be considered satisfied, each receiver $u$ decodes a message with a preference rank not exceeding $\eta_u$. 
\item Consider an $\mathcal{S}$. For each receiver node~$r_i \in W_1(\mathcal{S})$, let $b_i$ be the only node in $\mathcal{S}$ that is connected to $r_i$.
PrGrCov dictates that receiver~$i$ decodes $X_{b_i}$ from sub-code $\sum_{j \in \mathcal{S}}X_j$ generated by $\mathcal{S}$.\footnote{By design of the algorithm, each receiver decodes the message that $\mathcal{S}$ satisfies the receiver with. The algorithm disregards the possibility that the receiver may decode a more preferred message using a combination of coded messages from the entire code. We will address this points in Section~\ref{sec:further-improvement}.}
Define $M(\mathcal{S}) := \sum_{r_i \in W_1(\mathcal{S})} w(r_i,b_i) =\sum_{r_i \in W_1(\mathcal{S})}P_{i,b_i}$. $M(\mathcal{S})$ is thus the sum \textit{satisfaction ranks} of the receivers satisfied by $\mathcal{S}$. As the number of receivers satisfied by $\mathcal{S}$ is $|W_1(\mathcal{S})|$, $\frac{M(\mathcal{S})}{|W_1(\mathcal{S})|}$ is then the \emph{average} satisfaction rank of these receivers.
\item The function $f(\mathcal{S})$ is a weighted sum of (i) the number of receivers satisfied, and (ii) the average satisfaction rank of these receivers. The average satisfaction rank is given a negative weighting as a smaller rank is preferred.
\item Define $\eta_\text{max} := \max_{i \in [1:n]} \eta_i$. With this, $\alpha |W_1(\mathcal{S})| - (1 - \alpha) \frac{M (\mathcal{S}) } { |W_1(\mathcal{S})| } \geq -\eta_\text{max}$, regardless of $\alpha$.
$f(\mathcal{S})$ is then assigned a value of $-(\eta_\text{max}+1)$ when $\mathcal{S}$ satisfies no receiver, so that $\mathcal{S}$ that can satisfy one or more receivers is always preferred over $\mathcal{S}$ that satisfies none.
\item The parameter $\alpha \in [0,1]$ serves as a lever to weigh the two objectives: satisfying more receivers per sub-code and delivering messages with better preference ranks. In the next section, we will explain how adjusting $\alpha$ affects $\ell$ and $s$, and support the claim with simulation results.
Note, selecting $\alpha = 1$ and $\eta_i \geq \max_{j \in [1:m] \setminus H_i} P_{i,j}$ for all $i \in [1:n]$, we recover GrCov. 
\end{enumerate}

Our proposed Preferential Greedy Cover (PrGrCov) algorithm is outlined in
Algorithm~\ref{alg:pr-grcov}. The algorithm has time complexity $\mathcal{O}(n^2m^2)$.
See Appendix~\ref{app:complexity} for a discussion.


\subsection{Tuning parameters for PrGrCov}

PrGrCov takes in two tuning parameters: $\alpha$ and $\boldsymbol{\eta}$. Recall that the aim of the problem formulation is to find the Pareto boundary, that is, the optimal trade-off boundary between the code length and the satisfaction metric.

Parameter $\alpha$ attempts to balance these two opposing objective functions by weighing the following during the generation of each sub-code (which is a function of $\mathcal{S})$ (i) the number of receivers that $\mathcal{S}$ satisfies, $|W_1(\mathcal{S})|$, and (ii) the averaged satisfaction ranks of these receivers, $M(\mathcal{S})/|W_1(\mathcal{S})|$. 

\begin{figure}[t]
    \centering
    {\includegraphics[width=1.0\linewidth]{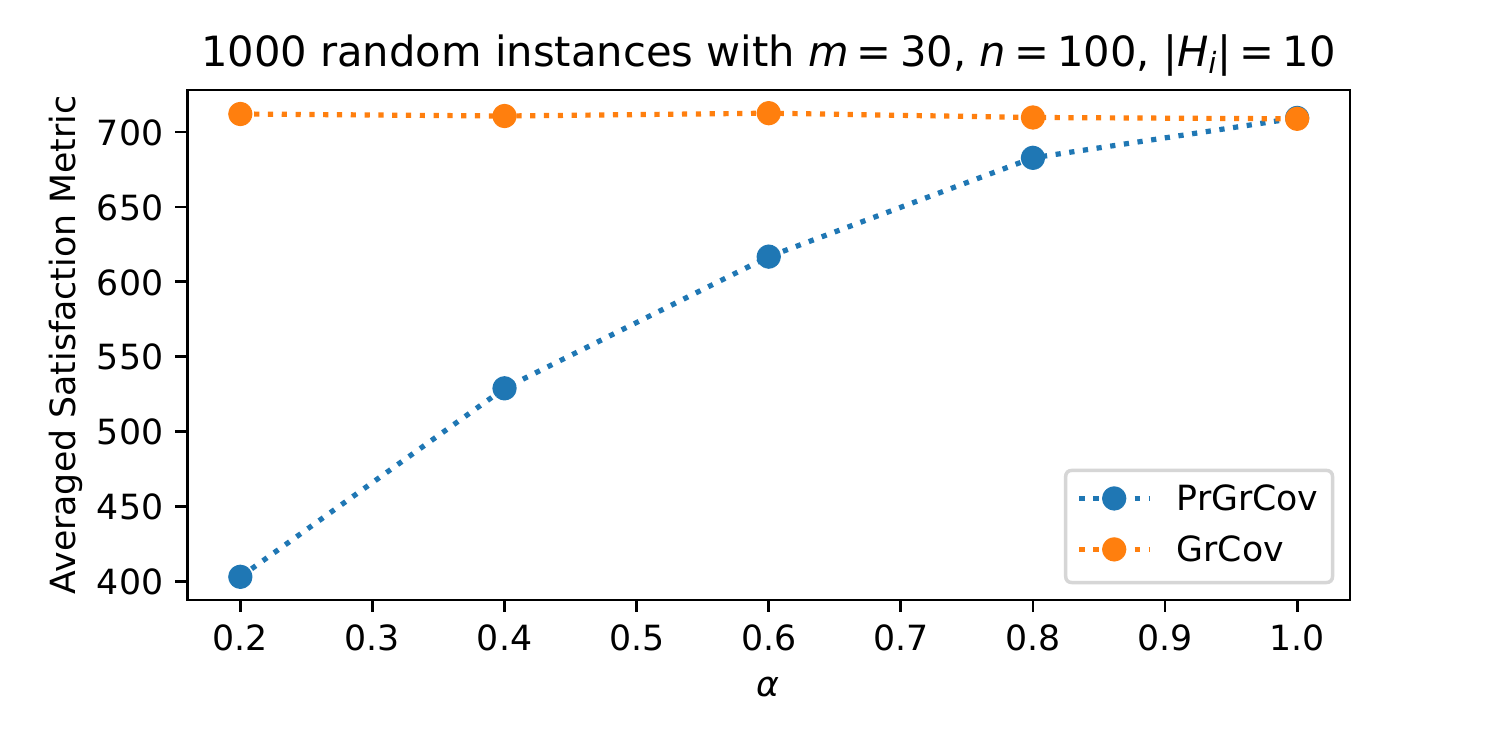}\vspace{-2ex}}
    {\includegraphics[width=1.0\linewidth]{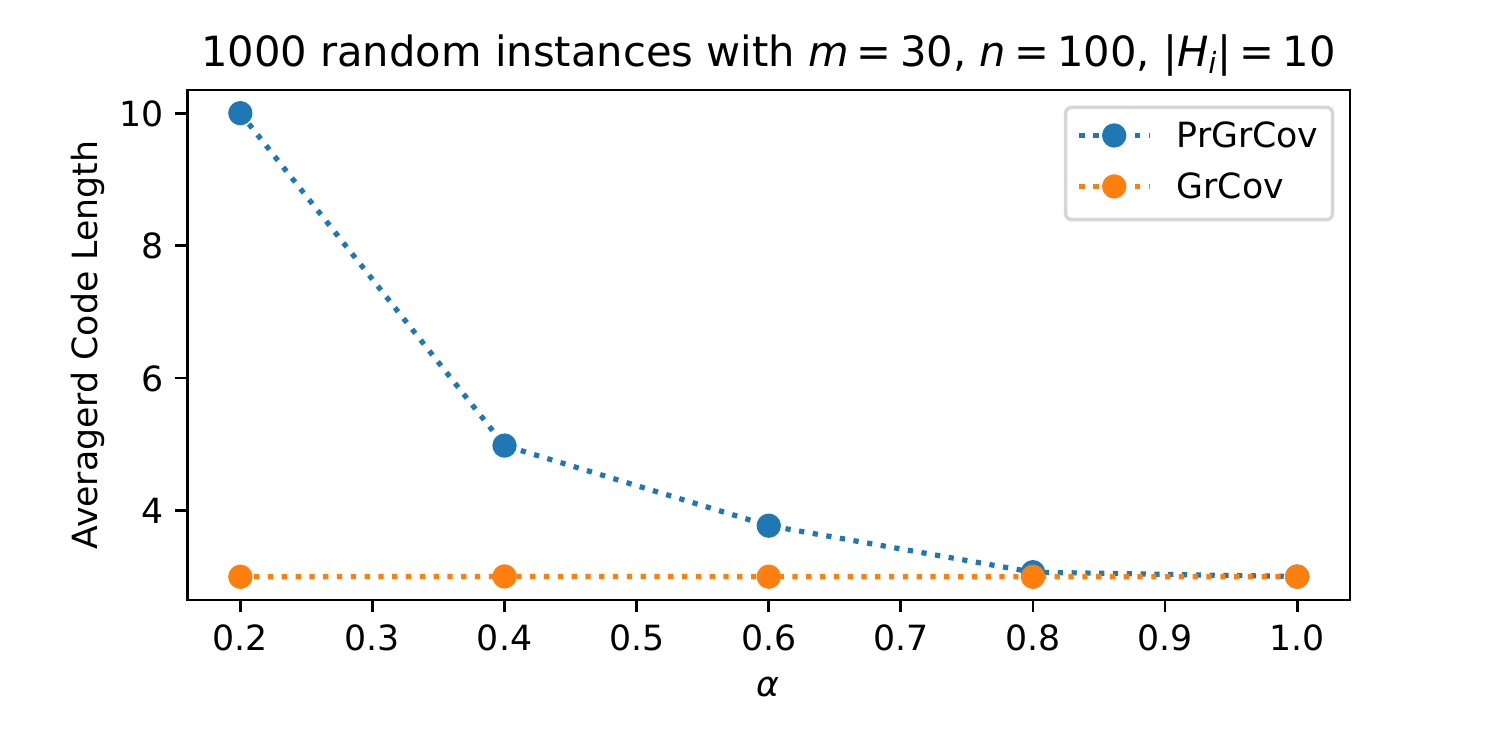}\vspace{-2ex}}
    \caption{The effect of $\alpha$ on satisfaction index and code length averaged over 1000 randomly generated PPICOD problems.}\label{fig:alpha}
\end{figure}

Using a larger $\alpha$, more receivers are satisfied per coded message, and fewer receivers are then left to be satisfied in subsequent iterations. Thus, the algorithm terminates earlier with a shorter code length.
Using a smaller $\alpha$, receivers are satisfied by messages with better satisfaction ranks. This results in codes with a better overall satisfaction metric.
Figure~\ref{fig:alpha} shows the effect of varying $\alpha$ on code length and satisfaction metric.

The inclusion of $\boldsymbol{\eta}$ is motivated by the way PrGrCov satisfies the receivers. Consider the case where the most preferred message for each receiver is ranked 1. The optimal satisfaction metric of $s^*=n$ is achievable by sending these messages uncoded. Let us set $\alpha=0$ to force the algorithm to only minimise the average satisfaction rank (and disregard the code length). However, suppose that in the first iteration of the algorithm, starting with $\mathcal{S} = \emptyset$, adding any message node $j$ to  $\mathcal{S}$ will satisfy some receiver~$i$ with a satisfaction rank of $P_{i,j} \in [2, \infty)$. After this iteration, the algorithm removes all receivers that the set $\mathcal{S}$ satisfies. In this case, the algorithm always ends with an overall satisfaction metric $s > n$. 

In order for the algorithm to disregard messages with poor preference rank for some receivers, albeit satisfied by the set of messages $\mathcal{S}$, the vector $\boldsymbol{\eta}$ is chosen such that only messages with preference ranks lower than the chosen values are considered for each receiver. 

For a fixed $\alpha$, using smaller $\boldsymbol{\eta}$ generally attains smaller $s$, usually at the expense of $\ell$. 
See Appendix~\ref{app:other-settings} for this behaviour.

By design, PrGrCov has the following characteristics:

\begin{proposition}
Choosing $\alpha = 1$ and $\eta_i \geq \max_{i \in [1:m]} P_{i,j}$ for all $i \in [1:n]$, Algorithm~\ref{alg:pr-grcov} specialises to the original GrCov algorithm for pliable index coding. 
\end{proposition}

\begin{proposition} \label{prop:lowest-score}
Choosing $\eta_i = \min_{j \in [1:m]} P_{i,j}$, for all $i \in [1:n]$, Algorithm~\ref{alg:pr-grcov} always produces a code that achieves $s^*=\min \big\{m, \sum_{i \in [1:n]} \min_{j \in [1:m]} P_{i,j}\big\}$. 
\end{proposition}

Although a code that achieves $s^*$ can be trivially constructed by sending either (i) $n$ uncoded messages $(X_{\text{any }k\in \arg \min_{j \in [1:m]} P_{i,j}}: i \in [1:n])$, each giving the best satisfaction for one receiver, or (ii) all $m$ messages, Proposition~\ref{prop:lowest-score} may achieve the same overall satisfaction metric with a shorter code length through coding. This is done by setting $0 < \alpha \leq 1$ so that over multiple choices of $\mathcal{S}$ in each iteration, one that satisfies more receivers is chosen. An example of such a scenario is provided in the next section.

\section{Results}


We consider a specific but sufficiently general structure of preferences $\boldsymbol{P}$, where the preference vector of each receiver~$i$ consists of consecutive integers starting from one in some order, that is, for each receiver~$i \in [1:n]$, 
\begin{equation}
    \{P_{i,j}: j \in [1:m]\setminus H_i\} = [1 : m - |H_i|]. \label{eq:increasing-rank}
\end{equation}

\begin{figure}[t]
    \centering
    \includegraphics[width=1.0\linewidth]{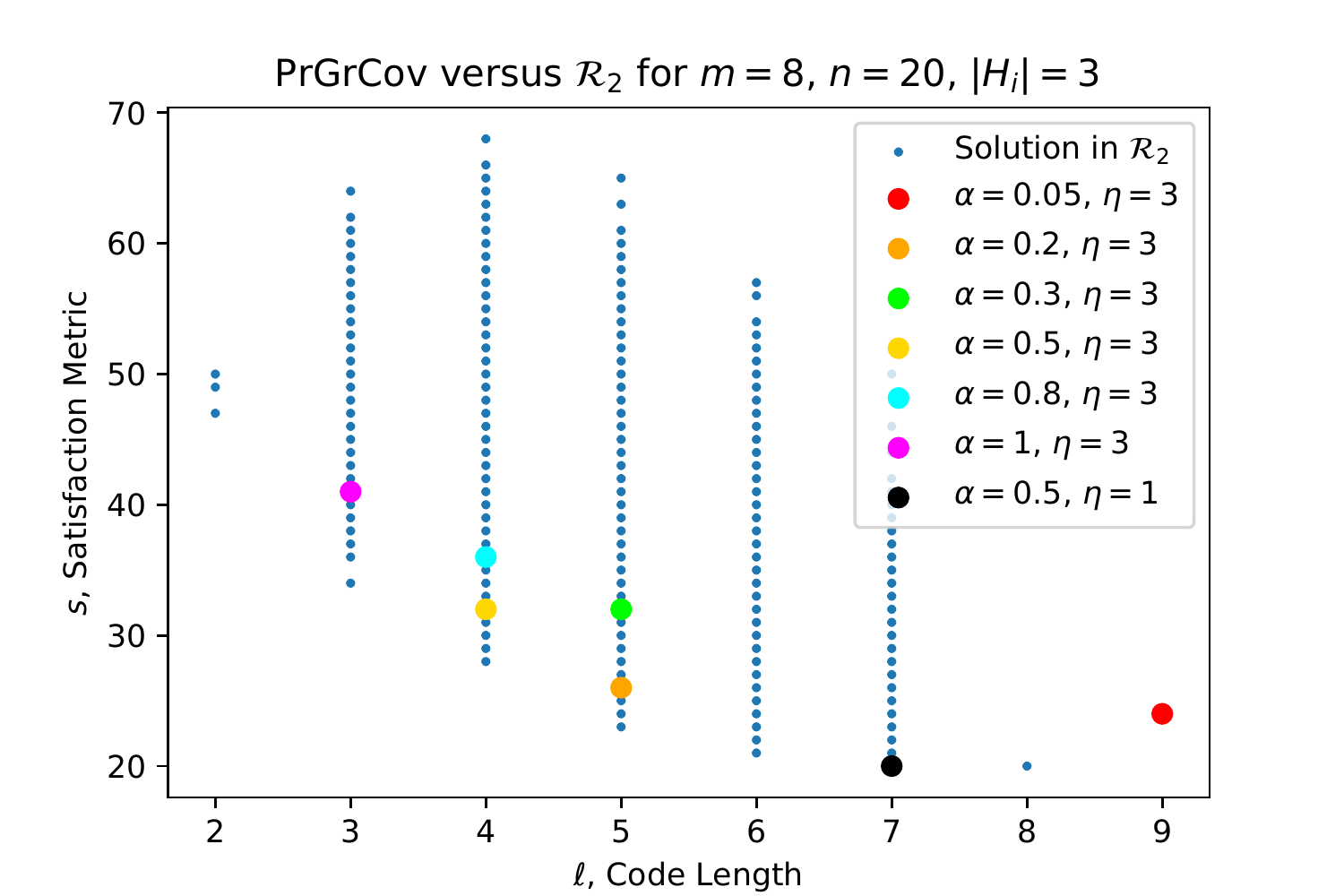}
    \caption{Code length--satisfaction pairs obtained by one instance of the PrGrCov algorithm with different $\alpha$ and $\eta=3$ versus the Pareto boundary of $\mathcal{R}_2$.}
    \label{fig:prgrcov-random-optimal}
\end{figure}

Figure~\ref{fig:prgrcov-random-optimal} shows the performance of PrGrCov for one specific PPICOD problem with $q=2$, $m=8$ messages, $n=20$ receivers, and $|H_i| =3$ for all $i \in [1:n]$. The side information and the preference ranks were randomly and independently generated for each receiver, subject to the aforementioned structure.

The performance of PrGrCov is compared with the Pareto boundary of $\mathcal{R}_2$ calculated using Method~2 in Section~\ref{sec:pareto}.

The tuning parameters are set to $\eta_i = \eta = 3$ for all $i \in [1:n]$ and $\alpha \in \{0.05, 0.2, 0.3, 0.5, 0.8, 1\}$.
Compared to $\mathcal{R}_2$ for $q=2$ (that is, binary linear codes), PrGrCov is capable of tracking the Pareto boundary well.

As pointed out earlier, if we set $\eta = 1$, PrGrCov is able to achieve the minimum satisfaction metric of $s^*=20$. In addition, for this specific PPICOD problem, it is also able to obtain the minimum code length $\ell^*=7$ at $s=20$, which outperforms the trivial transmission of $\min\{n,m\}$ uncoded messages.

We show in Appendix~\ref{app:other-settings} that PrGrCov can also perform well where groups of receivers have certain preference patterns.


\subsection{Further Improvement} \label{sec:further-improvement}

The code found by PrGrCov could be further processed to potentially improve the code length and the satisfaction metric:
\begin{itemize}
    \item The code is formed by transmitting one linear sum of messages, $\sum_{i \in \mathcal{S}} X_i$, for every $\mathcal{S}$ found during the iteration (lines 6--14 of the algorithm). As these sums may be linearly dependent, a shorter code length may be achieved by transmitting only linearly independent sums.
    \item The decoding choice of each receiver is determined by the set $\mathcal{S}$ that satisfies the receiver identified in the algorithm. The overall code could potentially allow the receiver to decode another message with a lower preference rank, and thus achieving a better overall satisfaction metric.
\end{itemize}

\bibliography{otherpublications,mypublications}

\clearpage

\appendices
\section{Proof of Proposition \ref{rate-cap}}\label{sec:proof:rate-cap}
\begin{IEEEproof}[Proof of Proposition \ref{rate-cap}]
Suppose that the satisfaction metric $s$ is achieved for some decoding choice $D$. The following three codes can also achieve the satisfaction metric $s$.
\begin{enumerate}
    \item $\mathsf E(X_{[1:m]}) = \left(X_{D(i)}  : i \in [1:n]\right)$, whose code length is $n$.
    \item $\mathsf E(X_{[1:m]}) = X_{[1:m]}$, whose code length is $m$.
    \item An $(m, m - \min_i |H_i|)$ MDS code for sufficiently large $q$. This code has a length of $m - \min_i |H_i|$ and allows all receivers to decode all $m$ messages.
\end{enumerate}
Taking the shorter code among them completes the proof.
\end{IEEEproof}

\section{Proof of Proposition~\ref{prop:r1*=r*}} \label{app:r1*=r*}
\begin{IEEEproof}[Proof of Proposition~\ref{prop:r1*=r*}]
    Clearly, $\mathcal{R}_1 \subseteq \mathcal{R}$.
    Consider any point $(\ell,s) \in \mathcal{R}^*$. The code that achieves the point must have a specific decoding choice $D$ that gives satisfaction metric $s$ according to \eqref{eq:global-satisfaction}. Step~b in Method~1 would have considered this particular decoding choice. Since Step~b finds the minimum code length for the associated index coding problem (that is, for the PPICOD problem with decoding choice $D$), it would have found a code that achieves $(\ell',s) \in \mathcal{R}_1$ where $\ell' \leq \ell$. 
    
    By definition of $\mathcal{R}^*$, $\mathcal{R}$ cannot contain any $(\ell'',s)$ with $\ell'' < \ell$. And since $\mathcal{R}_1 \subseteq \mathcal{R}$, we must have $\ell' = \ell$, and hence $(\ell,s) \in \mathcal{R}_1$. So, $\mathcal{R}^* \subseteq \mathcal{R}_1$.
    Since $\mathcal{R}^* \subseteq \mathcal{R}_1 \subseteq \mathcal{R}$, the proof is complete by invoking Lemma~\ref{lemma:subset-pareto}.
\end{IEEEproof}

\section{Proof of Lemma~\ref{lemma:subset-pareto}} \label{app:subset-pareto}
\begin{IEEEproof}[Proof of Lemma~\ref{lemma:subset-pareto}]
    All points in $\mathcal{R} \setminus \mathcal{R}^*$ have been excluded from the Pareto boundary of $\mathcal{R}$ as they are dominated by at least one point in $\mathcal{R}^*$. As $\mathcal{R}^* \subseteq \mathcal{R}_1 \subseteq \mathcal{R}$, all points in $\mathcal{R}_1 \setminus \mathcal{R}^*$ must be dominated by some point in $\mathcal{R}^*$, and hence would have been excluded from the Pareto boundary in $\mathcal{R}_1$ also. So, $\mathcal{R}_1^* \subseteq \mathcal{R}^*$. In addition, since all points in $\mathcal{R}^*$ are non-dominated in $\mathcal{R}$, they must also be non-dominated in $\mathcal{R}_1$. So, $\mathcal{R}^* \subseteq \mathcal{R}_1^*$. This gives $\mathcal{R}_1^* = \mathcal{R}^*$.
\end{IEEEproof}

\section{Proof of Proposition~\ref{prop:linear:pareto}} \label{app:linear:pareto}

\begin{IEEEproof}[Proof of Proposition~\ref{prop:linear:pareto}]
    Clearly, $\mathcal{R}_2 \subseteq \mathcal{R}$.
    Consider any point $(\ell,s) \in \mathcal{R}^*$. The point must be achieved by a linear encoding function of the form $\mathsf{E}(X_{[1:m]}) = \boldsymbol{A} \boldsymbol{X}$.

    Consider the decoding of receiver~$i$. We can express\footnote{For any matrix $\boldsymbol{A}$, let $\boldsymbol{A}_{H(i)}$ be the submatrix of $\boldsymbol{A}$ with only columns $H(i)$ retained, and $\boldsymbol{A}_{\setminus H(i)}$ be the submatrix of $\boldsymbol{A}$ with only columns $H(i)$ removed. With abuse of notation, for the column vector $\boldsymbol{X}$, let $\boldsymbol{X}_{H(i)}$ be the subvector of $\boldsymbol{X}$ with only rows $H(i)$ retained, and $\boldsymbol{X}_{\setminus H(i)}$ be similarly defined.} $\boldsymbol{Y} = \boldsymbol{A}_{\setminus H(i)} \boldsymbol{X}_{\setminus H(i)} + \boldsymbol{A}_{H(i)} \boldsymbol{X}_{H(i)}$.
    Since $D(i) \in [1:m] \setminus H_i$, receiver~$i$ can first perform $\boldsymbol{Y}' = \boldsymbol{Y} - \boldsymbol{A}_{H(i)} \boldsymbol{X}_{H(i)} = \boldsymbol{A}_{\setminus H(i)} \boldsymbol{X}_{\setminus H(i)}$ without affecting the decoding of $X_{D(i)}$, since $X_{D(i)}$ is independent of $X_{H(i)}$.

    In a consistent system of equations $\boldsymbol{M} \boldsymbol{x} = \boldsymbol{c}$, a variable $x_j$ (the $j$-th element of $\boldsymbol{x}$)  has a unique solution if and  only if a linear combination of the rows of $\boldsymbol{M}$ yields $\boldsymbol{e}_j$, the binary vector with a single one in the $j$-th component.

    So, there exists a matrix $\boldsymbol{B}$ such that $\boldsymbol{B} \boldsymbol{A}_{\setminus H(i)}$ contains a binary row vector of all zero except for the $D'(i)$-th position that corresponds to the position of $X_{D(i)}$ in $\boldsymbol{X}$. This means the reduced row echelon form of $\boldsymbol{A}_{\setminus H(i)}$ must have one row that has one on the $D'(i)$-th position and zero elsewhere.

    Note that $(\ell,s)$ lies on the Pareto boundary of $\mathcal{R}$. So for a given code $\boldsymbol{A}\boldsymbol{X}$ (which fixes $\ell$), if a receiver can decode two or more messages, the satisfaction metric of $s$ must have been achieved by the receiver decoding the message with the lowest preference rank.

    Repeating the above decoding argument for all the receivers, we show that Method~2 can achieve $(\ell,s)$, that is, $\mathcal{R}^* \subseteq \mathcal{R}_2$. Since $\mathcal{R}^* \subseteq\mathcal{R}_2 \subseteq \mathcal{R}$, the proof is complete by invoking Lemma~\ref{lemma:subset-pareto}.
\end{IEEEproof} 

\section{Time Complexity of PrGrCov} \label{app:complexity}

\begin{itemize}
    \item The \texttt{while} loop starting line~4 in Algorithm~\ref{alg:pr-grcov} iterates at most $n$ times because in every loop, $\mathsf{SAT}$ always increases in line~15 due to the definition of \eqref{eqn:fitness-func-improved}, where $|W_1(\mathcal{S})|>1$ is always preferred over $|W_1(\mathcal{S})|=0$.
    \begin{itemize}
        \item The \texttt{while} loop starting line~7 iterates at most $m$ times because at most $m$ messages can be added to $\mathcal{S}$ (starting from $\mathcal{S} = \emptyset$) and the condition in line~10 can be true for at most $m$ iterations of the \texttt{while} loop starting line~7.
        \begin{itemize}
            \item For line 9, $f(\mathcal{S} \cup \{j\})$, $\forall j \in [1:m] \setminus \mathcal{S}$, can be computed as follows:
            \begin{itemize}
                \item The computation of $W_1(\mathcal{S})$ requires checking of $n|\mathcal{S}|$ edges from each of the $n$ receivers to each message nodes in $\mathcal{S}$.
                \item The computation of $W_1(\mathcal{S} \cup \{j\})$ for one $j \in [1:m] \setminus \mathcal{S}$ can be done by modifying the results from $W_1(\mathcal{S})$ by checking at most $n$ edges from the receivers to message node~$j$. So, having $W_1(\mathcal{S})$, the computation of $W_1(\mathcal{S} \cup \{j\})$, $\forall j \in [1:m] \setminus \mathcal{S}$, involves checking at most $n(m - |\mathcal{S}|)$ edges.
                \item After that, the computation of $\frac{M(\mathcal{S} \cup \{j\})}{|W_1(\mathcal{S} \cup \{j\})|}$, $\forall j \in [1:m] \setminus \mathcal{S}$, has time complexity $\mathcal{O}(nm)$.
            \end{itemize}
            So, computing $f(\mathcal{S} \cup \{j\})$, $\forall j \in [1:m] \setminus \mathcal{S}$, has time complexity $\mathcal{O}(nm)$.
            \item The $\texttt{arg max}$ operation in line 9 requires at most $2m$ comparisons.
        \end{itemize}
        \item The edge pruning operation in line~17 deletes at most $mn$ edges.
    \end{itemize}
\end{itemize}
    So, the time complexity of PrGrCov is $\mathcal{O}(n^2m^2)$.

\section{More Simulation Results} 
\label{app:other-settings}

\begin{figure}[t]
    \centering
    \includegraphics[width=1.0\linewidth]{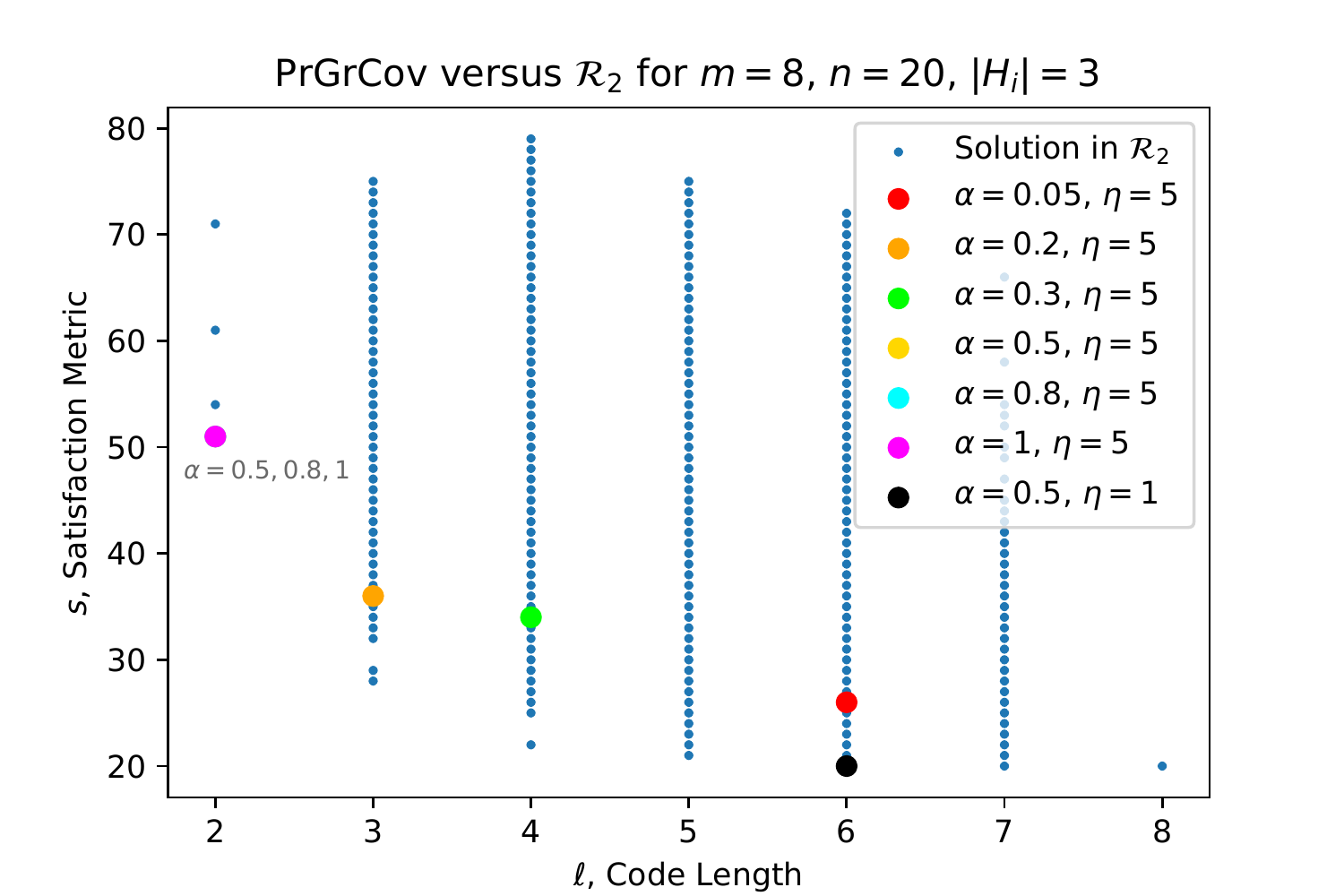}
    \caption{Code length--satisfaction pairs obtained by one instance of the PrGrCov algorithm with biased preferences, different $\alpha$ and $\eta=5$.}
    \label{fig:prgrcov-biased-1-optimal}
\end{figure}

\begin{figure}[t]
    \centering
    \includegraphics[width=1.0\linewidth]{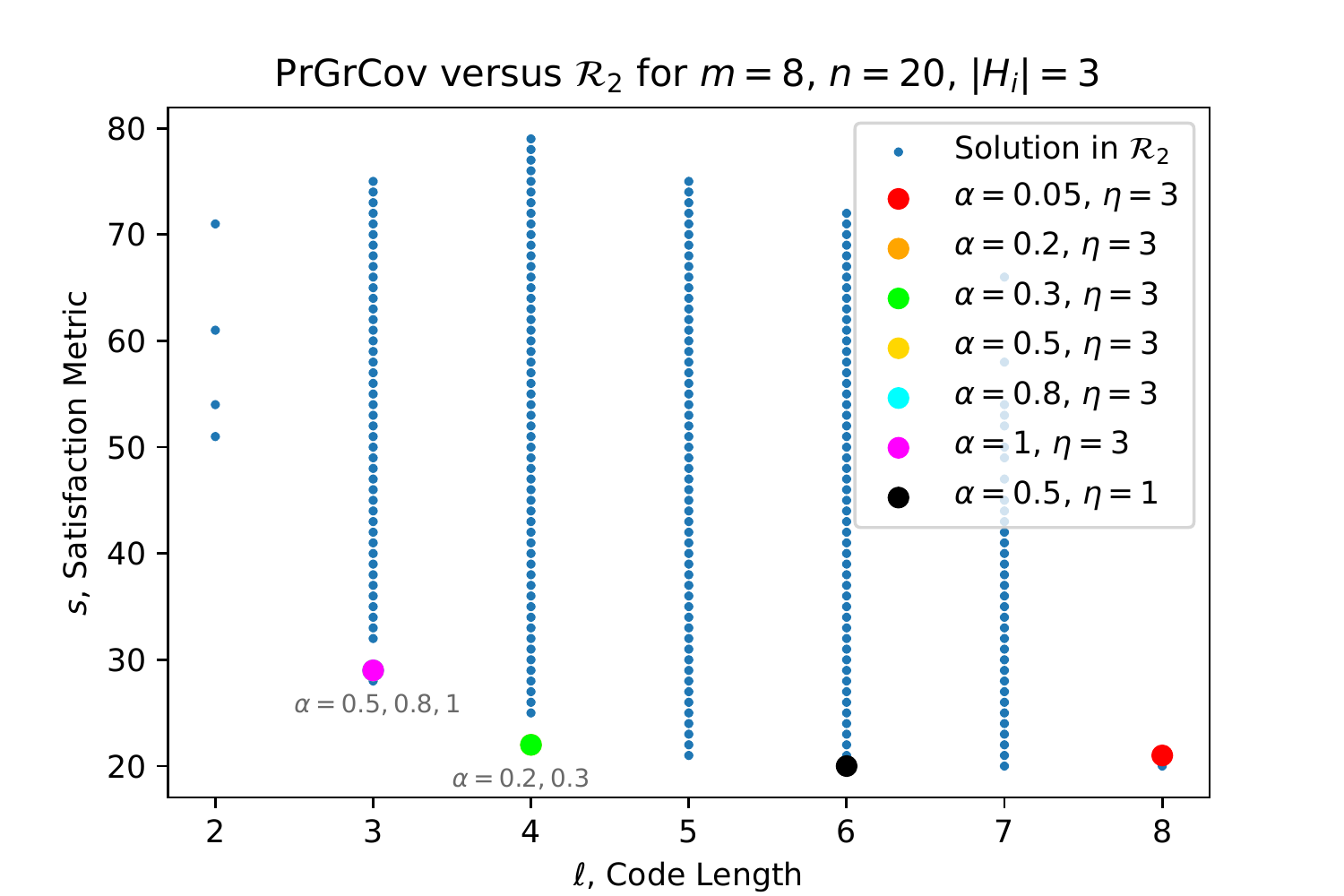}
    \caption{Code length--satisfaction pairs obtained by one instance of the PrGrCov algorithm with biased preferences, different $\alpha$ and $\eta=3$.}
    \label{fig:prgrcov-biased-2-optimal}
\end{figure}

The results depicted in Figure~\ref{fig:prgrcov-random-optimal} were obtained for a PPICOD problem where the preference ranks and side information sets for each receiver were randomly generated. 

Consider another setting where the side information sets are randomly generated but the preference ranks of the receivers follow a certain pattern. This mimics the scenarios where the receivers belong to different groups with particular group preferences.

Similar to the setting for Figure~\ref{fig:prgrcov-random-optimal}, we keep $q=2$, $m=8$ messages, $n=20$ receivers, $|H_i| =3$ for all $i \in [1:n]$, and~\eqref{eq:increasing-rank}.

The first group of receivers, receivers $k \in [1:10]$, prefer messages 1--4 and have the following preference pattern:
\begin{itemize}
    \item For any $i,j \in [1:m] \setminus H_k$, if $i < j$, then $P_{k,i} < P_{k,j}$.
\end{itemize}
An example preference vector is this: $$\begin{bmatrix}
\infty & 1 & 2 & \infty & 3 & \infty & 4 & 5
\end{bmatrix}.$$

The second group of receivers, receivers $k \in [11:20]$, prefer messages 5--8 and have the following preference pattern:
\begin{itemize}
    \item For any $i,j \in [1:m] \setminus H_k$, if $i + 3 \mod 8 < j + 3 \mod 8$, then $P_{k,i} < P_{k,j}$.
\end{itemize}
An example preference vector is this: $$\begin{bmatrix}
\infty & 3 & 4 & 5 & \infty & 1 & \infty & 2
\end{bmatrix}.$$

Figures~\ref{fig:prgrcov-biased-1-optimal} and \ref{fig:prgrcov-biased-2-optimal} show the performance of PrGrCov (with different $\alpha$ and $\eta$) compared to the Pareto boundary of $\mathcal{R}_2$ for one specific PPICOD problem with such group-wise biased preferences.

We note that
\begin{itemize}
    \item PrGrCov is capable of performing close to the Pareto boundary of $\mathcal{R}_2$ for group-wise preferences.
    \item Using a smaller $\eta$ generally results in a smaller $s$ and larger $\ell$.
\end{itemize}
\end{document}